\documentclass[11pt]{article}
\pdfoutput=1

\usepackage{jcappub}

\usepackage{amsfonts} 
\usepackage{amsmath} 

\usepackage{graphicx} 
\usepackage{float}
\usepackage[margin=1in]{geometry}
\usepackage{float}
\usepackage{tabularx}
\usepackage{array}
\usepackage{subfigure}
\usepackage{epstopdf}

\title{Local random potentials of high differentiability  to model the Landscape}

\author[a,b]{Thorsten Battefeld}
\author[c,d]{Chirag Modi}
\affiliation[a]{Department of Physics,
University of Minnesota,
1023 University Dr.,
Duluth, MN 55812, U.S.A.}
\affiliation[b]{Institute for Astrophysics, University of Goettingen, Friedrich Hund Platz 1, D-37077 Goettingen, Germany}
\affiliation[c]{Indian Institute of Technology- Bombay, Powai, Mumbai-40076, India}
\affiliation[d]{Department of Physics,University of California, Berkeley, CA 94720, U.S.A.}

\emailAdd{tbattefe@gmail.com, \\ modichirag@berkeley.edu}

\abstract{We generate random functions locally via a novel generalization of Dyson Brownian motion, such that the functions are in a desired differentiability class $C^k$, while ensuring that the Hessian is a member of the Gaussian orthogonal ensemble (other ensembles might be chosen if desired). Potentials in such higher differentiability classes ($k\geq 2$) are required/desirable to model string theoretical landscapes, for instance to compute cosmological perturbations  (e.g.,  $k=2$ for the power-spectrum) or to search for minima (e.g., suitable de Sitter vacua for our universe). Since potentials are created locally, numerical studies become feasible even if the dimension of field space is large ($D\sim 100$).
In addition to the theoretical prescription, we provide some numerical examples to highlight properties of such potentials; concrete cosmological applications will be discussed in companion publications. }

\arxivnumber{xxxx.xxxx}

\begin{document}
\maketitle
\flushbottom

\section{Introduction}
\label{sec:intro}

Random functions have many application in physics and mathematics, one of the best known ones is their use to describe disordered systems in solid state physics leading to Anderson localization (often Gaussian random potentials based on a truncated Fourier series are used). In this paper we derive, to our knowledge, new methods to generate random functions of high differentiability locally, while retaining a Hessian in the Gaussian orthogonal ensemble. Our motivation stems from our desire to study cosmological implications of certain landscapes in string theory, but we tried to make our results accessible to a wider audience by delegating cosmological applications to a separate publication. Readers not familiar with cosmology or string theory may simply skip the motivational paragraphs.

As alluded to, a recent application that motivated this work is the use of random functions to model certain landscapes in string theory \cite{Susskind:2003kw}. For example, the Denev-Douglas landscape \cite{Denef:2004cf} was modelled by a random potential in \cite{Marsh:2011aa,Bachlechner:2014rqa} (``Random Supergravities''), see also \cite{Chen:2011ac}. Since a top down approach yielding the full potential is virtually impossible in all but the simplest cases, random potentials are used to conduct numerical experiments and search for suitable vacua \cite{Chen:2011ac,Bachlechner:2012at}, investigate the feasibility of inflation \cite{Aazami:2005jf,Battefeld:2012qx,Yang:2012jf,BlancoPillado:2012cb} or compute (distributions of) observables\footnote{A concrete example of this approach is within the KKLMMT \cite{Kachru:2003sx}
brane inflation proposal \cite{Dvali:1998pa,Alexander:2001ks,Burgess:2001fx,Dvali:2001fw, Firouzjahi:2003zy,Burgess:2004kv,Buchel:2003qj,Iizuka:2004ct} with dynamical angular directions: certain random potentials are used to model the field's potential \cite{Agarwal:2011wm} so that the distribution of observables can be computed \cite{McAllister:2012am} (see also \cite{Dias:2012nf} whose discrepancies with \cite{McAllister:2012am} are explained in \cite{McAllister:2012am}).} \cite{Easther:2013rva,Westphal:2012up,Frazer:2011tg,McAllister:2012am,Dias:2012nf,Pedro:2013nda,Battefeld:2013xwa}, see also \cite{Tye:2008ef,Linde:2007jn,Sumitomo:2012cf,Sumitomo:2012vx,Duplessis:2012nb} for related work (for recent reviews of inflation see \cite{Bassett:2005xm,Baumann:2009ds} and for model building in string theory see \cite{Baumann:2014nda}). Naturally, the closer random potentials model the actual landscape of interest, the more reliable predictions become. Thus, our interest is to prescribe certain generic properties, such as the overall hilliness, the properties of the Hessian at well separated points etc., whenever a random potential is generated. For example, in random supergravities, the Hessian is a mix of Wishart and Wigner matrices \cite{Marsh:2011aa}. A complementary analytic tool is random matrix theory (see \cite{Mehta:1991,Rao:2005} for a textbook introduction) which can often be used in conjunction with numerical experiments due to the feature of universality \cite{Deift:2006,Kuijlaars:2011,Bai,Soshnikov:2002,Schenker:2005}. Most work in recent years relied on potentials constructed globally via truncated Fourier series \cite{Tegmark:2004qd,Aazami:2005jf,Frazer:2011tg,Frazer:2011br,Battefeld:2012qx,Battefeld:2013xwa,Pedro:2013nda}, a subclass of which are Gaussian random potentials. However, this approach has the disadvantage of being computationally intensive as the dimensionality $D$ of field space increases, since the number of random parameters increases as $\#^D$. Thus, for a description of the hundreds of fields on generic string theoretical landscapes \cite{Susskind:2003kw}, this approach becomes useless.

Fortunately, many questions of interest, for instance the likelihood of inflation, the probability of encountering a minimum or the values of observables such as the scalar spectral index (the slope of the observed power-spectrum), only require knowledge of the scalar fields' potential (the landscape) in the vicinity of the trajectory taken by the fields. This motivated Marsch et al.~\cite{Marsh:2013qca} to construct the potential locally by employing Dyson Brownian motion \cite{Dyson:62} (DBM), greatly reducing the cost of numerical experiments ($\propto D^{\#}$) to the point where $\mathcal{O}(100)$ fields can be treated on a notebook with Mathematica. 

To generate a potential via DBM, one stitches together patches wherein the potential is given by a Taylor series, truncated at second order. After a prescribed step, for instance a set distance along an inflationary trajectory, random components are added to the Hessian. In DBM, these random components conform to a Gaussian distribution with prescribed mean and variance, so that the Hessian is a member of the Gaussian orthogonal ensemble.

While efficient, this procedure has a serious drawback: after each step the eigenvalues of the Hessian, i.e.~the masses of fields, jump. Such potentials are ill suited to study cosmological perturbations, since artefacts arise in correlation functions. For example, even a single jump in the mass of one of the fields causes a prominent ringing pattern in the power-spectrum \cite{Joy:2007na,Joy:2008qd,Battefeld:2014aea} (i.e.~the two point correlation function). Higher order correlation functions, commonly lumped together under the name non-Gaussianities, are affected even more. Such jumps in the Hessian can also hinder the search for minima: whenever a minimum is approached, the Hessian starts to dominate the evolution; due to the random jumps, the trajectory is bounced around preventing a smooth approach to the minimum. To a lesser degree the steps can also reduce the probability of finding regions that are sufficiently flat for inflation. It is possible to reduce these artefacts by decreasing the step size; however, this brute force approach reduces the computational advantage of DBM.

Thus, in this paper, we generalize the method by Dyson to generate potentials in any desired differentiability class: we delegate perturbations not to the Hessian, but to higher derivative tensors, while retaining a Hessian in the Gaussian orthogonal ensemble (other distributions may be chosen if desired); for example, if $V\in C^2$ is desired, random Gaussian perturbations are added to the tensor of third derivatives $\partial^3 V/(\partial \phi_a\partial\phi_b\partial\phi_c)$. 

After a brief review of the method to generate potentials $V\in C^1$ via DBM in Sec.~\ref{sec:review1}, we provide two distinct methods to generate  $V\in C^2$, both of which yield the same statistical properties of the Hessian. Additional freedom is present, since the number of random variables exceed the number of conditions stemming from the prescribed statistical properties of the Hessian. The first method provides potentials that are ``smoothest'' in the sense that a maximal number of random variables is set to zero, Sec.~\ref{Sec:con}. The second methods adds perturbations primarily in the directions set by the eigenvalues of the Hessian, Sec.~\ref{Sec:cond2}. We compare both methods in Sec.~\ref{Sec:finaldiscussion} and find them to be qualitatively indistinguishable and free of artefacts. We plan to use these potentials for cosmological applications in a forthcoming publication, where we also intend to test how sensitive observables are to the chosen method -- naturally, any dependence would dramatically reduce the predictiveness and thus reduce the usefulness of such random potentials to model concrete landscapes in string theory. However, if observables are independent of the methods, we have the opportunity to compute generic predictions for whole classes of string theoretical landscapes, as opposed to investigating inflationary models on a case by case basis.

While potentials $V\in C^2$ are sufficient to compute the power-spectrum, they should not be used for higher order correlations functions. For example, if one wants to compute the bi-spectrum, one needs to be able take three derivatives of the potential, i.e.~$V\in C^3$ is needed. We therefore provide a generalization of the first method to create potentials in any desired differentiability class ($V\in C^k$ with $k\in \mathbb N$) in Sec.~\ref{Sec:potinCk}. For $k=3$, we find that spurious oscillations arise in the evolution of the Hessian's eigenvalues, Sec.~\ref{Sec:finaldiscussion}. These oscillations are caused by truncating the Taylor series at higher order and can't be avoided within the current framework; nevertheless, their amplitude, and thus their effect on observables, can be reduced to any desired level by decreasing the step length. While not an ideal solution, such potentials still improve on potentials of lower differentiability (one can't compute the bi-spectrum at all if $V\in C^1$). We leave future applications of such potentials as well as improvements to the methods put forth in this article to future work.\\

We would like to reiterate that the methods introduced in this study are independent of the applications outlined above.

\section{Creating random potentials along a trajectory}
\label{sec:review1}
\subsection{Motivation and goals}
Inflationary observables depend only on properties of the potential in the vicinity of the trajectory, which motivated Marsh et.al.~\cite{Marsh:2013qca} to develop a computationally economical approach to generate random potentials locally by defining random functions around a path  $\Gamma$ in field space\footnote{To achieve a single valued potential, $\Gamma$ should not be self intersecting. While self intersecting paths are common in two dimensional field spaces, they become exceedingly rare for larger $D$ (if $\Gamma$ is a random walk, it's fractal dimension is $2$ \cite{Rudnick:87,Rudnick:2004}; a random walk is a good approximation if $\Gamma$ is the solution to the field equations on a random potential during inflation). }: for any $\Gamma$, given the value of the potential $V$, gradient $V\rq{}$ and Hessian $V\rq{}\rq{} \equiv \mathcal H$ at a point $p_0$, the values of the potential and the gradient vector at a nearby point $p_1$ can be obtained to leading order by means of a Taylor expansion. To construct a random potential, the Taylor expansion is truncated and the Hessian matrix at $p_1$ is altered by adding a random matrix $\delta \mathcal{H}$ to the Hessian at $p_0$,
\begin{equation}
\mathcal H(p_1) = \mathcal H(p_0) + \delta \mathcal H\,.
\end{equation}
By repeating this process along the path $\Gamma$, a continuously differentiable, random potential, i.e.~$V \in C^1$, can be obtained.

The distribution of the Hessian matrix at well-separated points (i.e.~separated by several units of a characteristic correlation length $\Lambda_h$) can be restricted to any desired distribution; if Wigner's Gaussian Orthogonal Ensemble (GOE) is chosen, as in \cite{Marsh:2013qca}, the elements of the Hessian undergo Dyson Brownian motion (DBM) \cite{Dyson:62}. 
As a consequence of statistical rotational invariance, Hessian matrices associated with well-separated points constitute a random sample of the statistical ensemble, which is invariant under orthogonal transformations. Further, if the field space is $D$-dimensional, the $D(D + 1)/2$ entries of the Hessian matrix $\mathcal{H}$ are statistically independent\footnote{Another advantage of potentials generated via DBM as opposed to a trunctaed Fourier series is this statistical independence, which truncated Fourier potentials lack, see \cite{Marsh:2013qca} for a comparison.}. While the choice of the GOE is the simplest one, it is by no means unique; in concrete applications, e.g.~to construct potentials obeying prescribed properties of a landscape, the rules of constructing the potential have to be adjusted accordingly. 

We review Dyson Brownian motion in more detail in the next section, before generalizing the prescription. We are particularly interested in two aspects:

\begin{enumerate}
\item Generate potentials $V\in C^k$ with $k\geq 2$ (Sec.~\ref{Sec:VinC2} and Sec.~\ref{Sec:potinCk}), which is needed to compute correlations functions (e.g., $k=2$ for the power-spectrum, $k=3$ for the bi-spectrum etc.) if artefacts are to be avoided. In addition, $V\in C^2$ is desirable for searches of extrema (if a critical point is approached, the jumps in the Hessian in ordinary DBM hinders a localization/identification of extrema). 
\item Incorporate a soft upper and lower bound on the values of the potential, as in \cite{Battefeld:2012qx}; such a bound is necessary if the potential is used to model a low energy, effective potential.
\end{enumerate}

In this article, we focus on the first point, the generation of random functions $V\in C^k$, which provides the foundation for concrete applications, such as the computation of cosmological perturbations, or further refinements, such as the incorporation of bounds mentioned in point 2. The latter topics are the subject of companion publications (in preparation).

\subsection{Review: Dyson Brownian motion potentials \label{sec:DBMReview}}
 Dyson Brownian Motion is a canonical -- but not unique -- choice of rules to govern the stochastic evolution of the Hessian matrix that gives rise to independent GOE Wigner matrices at well-separated points. To this end, the Hessian needs to be perturbed according to (see \cite{Dyson:62,Marsh:2013qca}, whose results we  summarize here)
\begin{eqnarray}
\delta \mathcal{H}_{ab}=\delta A_{ab} -\mathcal{H}_{ab}\frac{\delta s}{\Lambda_h}\,,
\end{eqnarray}
where $\delta A_{ab}$ are $D(D + 1)/2$ zero-mean stochastic variables and the term $\propto -\mathcal{H}_{ab}$ is the uniquely determined restoring force ensuring that the distribution of the entries of the Hessian remains finite and obeys the GOE. This restoring force does not imply the boundedness of the potential. The variable $s$ represents the field space path length along the trajectory $\Gamma$ and $\delta s$ is the length of an individual step along $\Gamma$. $\Lambda_h$ can be interpreted as a horizontal correlation length. To achieve Dyson Brownian motion of $\mathcal{H}_{ab}$,  the first two moments of $\delta \mathcal{H}_{ab}$ need to satisfy \cite{Marsh:2013qca}

\begin{eqnarray}
 \langle \delta \mathcal{H}_{ab}|_{p_1}\rangle &=& -\mathcal{H}_{ab}|_{p_0} \frac{\delta s}{\Lambda_h}\,, \label{wignercondition1} \\
\langle (\delta \mathcal{H}_{ab})^2\rangle &=& (1+\delta_{ab}) \frac{\delta s}{\Lambda_h}\sigma^2\,, \label{wignercondition2} 
\end{eqnarray}
 where $\sigma$ represents the standard deviation of the corresponding Wigner ensemble.

To implement the above prescription, consider a $D$-dimensional field space with fields $\phi_a$, $a=1\dots D$ and a potential $V$. We would like to model $V$ as a random one given a suitable starting position. The potential in the vicinity of the starting point $p_0$ can be expanded as 
\begin{eqnarray}
V= \Lambda_v^4 \sqrt{D}\left[v_0 + v_a\tilde{\phi}^a+ \frac{1}{2}v_{ab}\tilde{\phi}^a\tilde{\phi}^b+\dots\right]\,, \label{taylorpotential}
\end{eqnarray}
where $\Lambda_v$ (mass dimension one) sets the vertical scale,
and $\tilde{\phi}^a \equiv \phi^a /\Lambda_h$ are rescaled, dimensionless fields. 
 The normalization factor $\sqrt{D}$ is introduced to simplify subsequent expressions. We use Einstein's summation convention over field indices and consider a flat field space metric if not stated otherwise. 

If we take the truncated Taylor expansion in (\ref{taylorpotential}) at $p_0$, the potential at an adjacent point $p_1$ close to $p_0$, i.e.~with local coordinates $\delta \tilde{\phi}^a$ satisfying $\|\delta \tilde{\phi}^a\| \ll 1$ where $\|\dots \|$ is the Cartesian norm, can be written as 
\begin{eqnarray}
v_0\!\mid_{p_1} &=& v_0\!\mid_{p_0} + v_a\!\mid_{p_0} \delta \tilde{\phi}^a + \frac{1}{2}v_{ab}\!\mid_{p_0} \delta \tilde{\phi}^a \delta\tilde{\phi}^b+\dots \\
 v_a\!\mid_{p_1} &=& v_a\!\mid_{p_0} + v_{ab}\!\mid_{p_0} \delta \tilde{\phi}^b+\dots\\
 v_{ab}\!\mid_{p_1} &=&v_{ab}\!\mid_{p_0}+\dots
\end{eqnarray}
To generate a random potential, one can truncate the series expansion at second order and set
\begin{eqnarray}
 v_{ab}\!\mid_{p_1} &=&v_{ab}\!\mid_{p_0}+\delta v_{ab}\!\mid_{p_0}\,,
\end{eqnarray}
 where $\delta v_{ab}\!\mid_{p_0}$ are taken to be elements of a random matrix\footnote{If the full potential were known, one would identify $\delta v_{ab}=v_{abc}\!\mid_{p_0} \delta \tilde{\phi}^c$.}. Repeated application over successive points $p_n$ along $\Gamma$ results in a random, piecewise patched together potential $V\in C^1$.
{\em Dyson Brownian motion random potentials} are therefore defined by imposing
\begin{eqnarray}
 \langle \delta v_{ab}\!\mid_{p_n}\rangle &=& -v_{ab}\!\mid_{p_{n-1}} \frac{\|\delta \phi^a\|}{\Lambda_h}\,,\label{condition1}\\
 \langle (\delta v_{ab}\!\mid_{p_n})^2\rangle &=& (1+\delta_{ab}) \frac{\|\delta \phi^a\|}{\Lambda_h}\sigma^2\label{condition2}
 \end{eqnarray}
 for the mean and second moment of the added random components $\delta v_{ab}$.  
 Since $O(v_{ab}) = 1/ \sqrt D$, the magnitude of a typical eigenvalue of $v_{ab}$ is of order one\footnote{The factor of $1/\sqrt{D}$, constitutes an overall normalization we used to remain consistent with \cite{Marsh:2013qca}. This choice, i.e.~a random matrix with $v_{ab} \sim \mathcal{O}(1/\sqrt{D})$, leads to eigenvalues of order one, a well known results in random matrix theory; this choice is convenient, since no small/large numbers appear numerically if D is increased in our Mathematica code.}. Thus, $v_0$ and $v_a$ both receive contributions of order $1/\sqrt D$ over a correlation length. Hence, for potentials uncorrelated over distances $\mathcal{O}(\Lambda_h)$, it is natural to take $O(v_0) = O(v_a) = O(v_{ab}) = 1/\sqrt {D}$, which in turn explains the overall normalization factor of $\sqrt{D}$ in (\ref{taylorpotential}).
 
\subsection{Eigenvalue relaxation \label{sec:eigenvaluerelaxation}}

The probability distribution of a matrix' eigenvalues in the unfluctuated GOE (i.e. the stationary distribution) is given by Wigners semi-circle law. If a Matrix is initialized in a fluctuated state, for example such that the smallest negative eigenvalue is close to zero, its eigenvalues relax to the stationary distribution as Dyson Brownian motion proceeds. The relevant correlation length is given by $\Lambda_h$. As a consequence, it is unlikely that a shallow patch  remains  flat for $s=\Delta{\boldsymbol{ \phi}}\gg \Lambda_h$ \footnote{For example, such an initially flat configuration would be chosen if inflation near a saddle point is to be investigated \cite{Marsh:2013qca}; evidently, if $\Lambda_h$ is small, long bursts of inflation are unlikely. This result can be quantified via random matrix theory, see e.g.~\cite{Aazami:2005jf,Battefeld:2012qx}.}.

To be concrete, one can show \cite{Marsh:2013qca} that the expectation value of the smallest eigenvalue $\lambda_{\mathrm{min}}$ of a matrix $\mathcal{H}$ undergoing DBM satisfies
\begin{eqnarray}
\label{lowerboundrelaxation} \langle\lambda_{\mathrm{min}}[\mathcal{H}(s)] \rangle
  \geq q \lambda_{\mathrm{min}}[\mathcal{H}^0]-2\sqrt{1-q^2}\,, 
\end{eqnarray}
where $\mathcal{H}^0$ is the initial matrix at $s=0$ and  $q\equiv \exp(-s/\Lambda_h)$. Here, a normalization such that eigenvalues lie between $-2$ and $2$ was chosen. Due to the exponential suppression in $q$, the initial eigenvalue $\lambda_{\mathrm{min}}[\mathcal{H}^0]$ is forgotten after a few $\Lambda_h$ are traversed.

\begin{figure}[H]
\centering
\subfigure[ $V\in C^1$ via Dyson Brownian Motion]
{\includegraphics[width=0.45\linewidth]{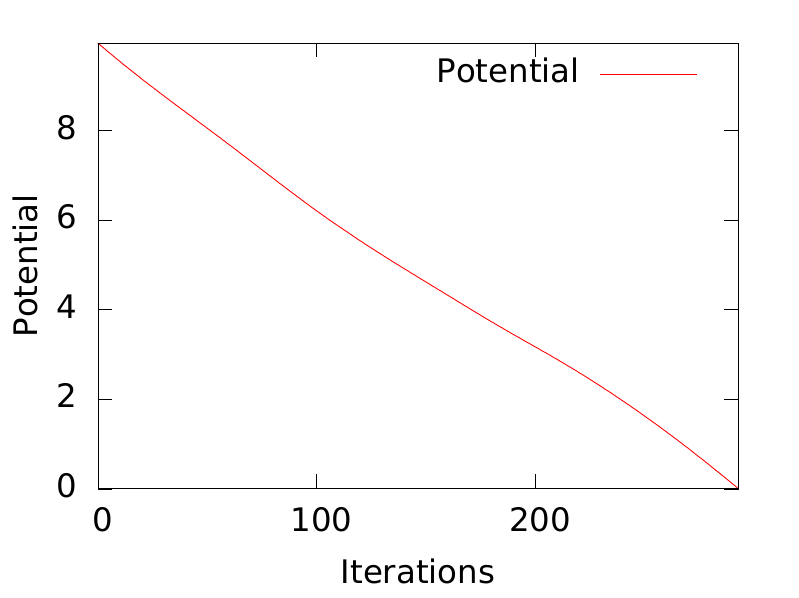}
 }
\subfigure[ $V\in C^1$ - Eigenvalue Relaxation.]
{\includegraphics[width=0.45\linewidth]{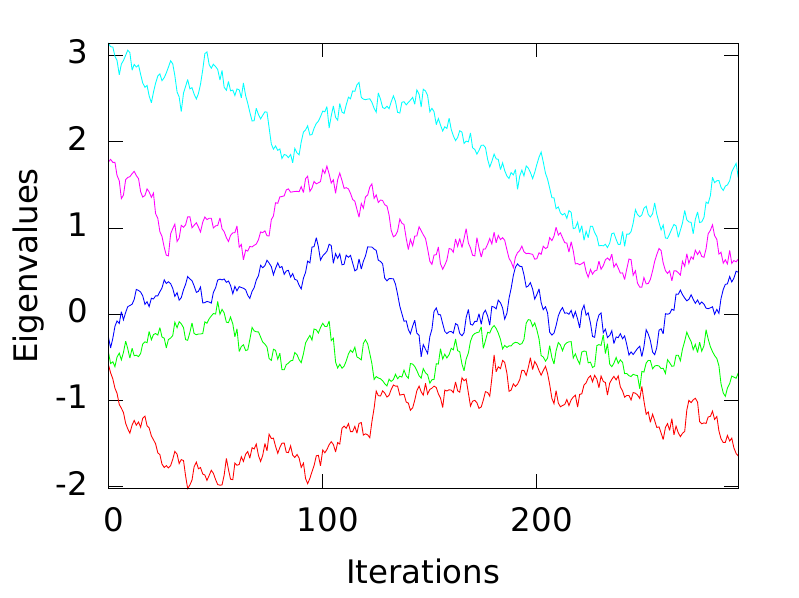}
}
\caption{ Potential $V\in C^1$ created via Dyson Brownian motion (a) and eigenvalue relaxation (b) for $D=5$ fields. A normalization is chosen such that Eigenvalues lie on average in the interval from $[-2\dots 2]$. Here and in all subsequent figures we chose $\Lambda_h=0.1$,  $\Lambda_v=1$, $\delta s= \Lambda_h/100$ and $\sigma^2=2/D$. Eigenvalues relax to the stationary distribution after about $\Lambda_h/\delta s=100$ iterations. \label{Fig:relaxationMarsch}}
\end{figure}

We demonstrate this eigenvalue relaxation and recover the results of Marsh et al.~\cite{Marsh:2013qca} in Fig.~\ref{Fig:relaxationMarsch}, where we plot a random potential $V\in C^1$ created via Dyson Brownian motion (using (\ref{taylorpotential}) and perturbations of the form (\ref{condition1}) and (\ref{condition2})) next to the corresponding eigenvalues of the Hessian. Evidently, eigenvalues relax to the stationary distribution after about $\Lambda_h/\delta s\sim 100$ iterations or steps.
Here and in all subsequent figures we take
 \begin{eqnarray}
M_{P}=\frac{1}{\sqrt{8\pi G}}\equiv 1
\end{eqnarray} 
 $\Lambda_v=M_{P}=1$, $\Lambda_h=0.1M_P=0.1$, $\delta s=\Lambda_h/100$ and $\sigma^2=2/D$ 
  if not stated otherwise. While the choices of $\Lambda_h$ and $\Lambda_v$ have implications for concrete applications\footnote{For instance, in a cosmological setting $\Lambda_v$ sets the energy scale in the Einstein equations and $\Lambda_h$ the smoothness of the potential and thus determines whether or not inflation is common.}, they merely correspond to a rescaling here. Thus, without loss of generality, we can keep them fixed. Further, as long as the step length is small enough ($\delta s\ll \Lambda_h$ has to hold) it should not have any impact on applications; how small it has to be depends on the type of application. We come back to this point in Sec.~\ref{Sec:finaldiscussion}. In the meantime, $\delta s=\Lambda_h/100$ is small enough for our  purposes, while big enough to enable fast computations. $\sigma$ controls the strength of the perturbations and thus affects the overall hilliness of $V$. We chose $\sigma^2=2/D$ in line with \cite{Marsh:2013qca}, to enable direct comparisons. To test our code, we recovered numerically the results of \cite{Marsh:2013qca}, which we omit here for reasons of brevity.
 
The path $\Gamma$ we choose is given by following the slope of the potential. In a cosmological setting, one is commonly interested in solving the field equations in conjunction with the Friedmann equations, but following the steepest descent  is faster and suffices for our purposes \footnote{During slow roll inflation, the trajectory  follows the direction of steepest descent to some degree, but differences can be crucial to address questions pertaining to observables such as the power-spectrum or the bi-spectrum, see e.g.~\cite{Tolley:2009fg,Achucarro:2010jv,Achucarro:2010da,Achucarro:2012sm,Achucarro:2012yr,Burgess:2012dz,Achucarro:2012fd,Achucarro:2014msa} for an effective field treatment and \cite{Gao:2012uq,Gao:2013ota,Noumi:2013cfa,Konieczka:2014zja} for numerical studies. However, the choice of $\Gamma$ has no bearing on the methods of generating the potential discussed in this article as long as $\Gamma$ is not self intersecting.}. 

While the observation of eigenvalue relaxation does not serve as a strong test for the ensemble to be the GOE (the same results hold true for other symmetric distributions with zero mean and finite variance for off diagonal elements), it provides a simple, necessary consistency check.
In the following, we use such plots as benchmarks: as we construct potentials in a higher differentiability class, which is achieved by perturbing a higher order derivative tensor instead of the Hessian while retaining the statistical properties of the Hessian, plots such as the ones above should remain qualitatively unchanged.

\section{Extending Dyson Brownian motion to generate random potentials $V\in C^2$ \label{Sec:VinC2}}
The perturbations of the Hessian in the aforementioned procedure yield a potential in $C^1$. If one wishes to study inflationary cosmology on random potentials, one is commonly interested in the evolution of cosmological perturbations to compute correlation functions of the gauge invariant curvature fluctuation $\zeta$. The latter can be measured by observations of the cosmic microwave background radiation (CMBR) or large scale structure surveys. Of particular interest are the two-point function (power-spectrum) and the three-point function (a measure of non-Gaussianities). Since $\zeta$ is related to fluctuations in the inflatons at horizon crossing, see \cite{Bassett:2005xm,Baumann:2009ds} for reviews, the correlation functions of $\zeta$ probe the properties of the inflationary potential around sixty e-folds before the end of inflation. 

The $n$'th correlation function is sensitive to the $n$'th derivative of the potential -- for instance, at the level of the slow roll approximation the second derivative of the potential enters the scalar spectral index. Thus, a discontinuity in the $2$'nd derivative of the potential, as induced by Dyson Brownian motion, leads to artefacts already at the level of the power-spectrum: as shown in \cite{Joy:2007na,Joy:2008qd,Battefeld:2014aea} such jumps lead to extended oscillations in the power-spectrum; higher order functions are affected as well.  Keeping $\delta s \ll 1$ and thus $\delta \mathcal{H}_{ab}$ sufficiently small, washes out these artefacts, since they are superimposed on top of each other as in \cite{Battefeld:2010vr}. However, such a brute force approach is computationally intensive, offsetting the main advantage of DMB, and potentially not entirely free of artefacts. Thus, it is desirable to have access to a random potential in the class $C^k$ if the $k$'th correlation function is to be computed.

Even if one is not interested in cosmological  perturbations but merely properties of the inflationary background trajectory, it is advisable to have  $k\geq 2$: for example, one might be interested in the final vacuum reached after inflation, as in \cite{Battefeld:2012qx}. To this end one needs to identify the presence of a local minimum. However, if the necessarily shallow region near the minimum is approached the gradient approaches zero; as a consequence, the background dynamics become dominated by the Hessian, not the gradient. We expect the sudden steps in $\mathcal{H}$ to hinder the proper identification of a minimum, since the artificial jumps prevent a smooth approach to it. This expectation can be seen numerically, and we plan to elaborate on this point in future work. To a lesser degree we expect this effect  to arise during slow roll inflation as well, particularly if inflation is driven near a saddle point or maximum.  


In this section we generate a random potential $V\in C^2$, such that the elements in the Hessian still obey the GOE, by adding the random fluctuations to the tensor of third derivatives instead of the Hessian. Such a potential is sufficient to discuss background dynamics, including questions pertaining to the final resting place, as well as the power-spectrum. A generalization to $V\in C^k$ is given in Sec.~\ref{Sec:potinCk}.

\subsection{A potential to third order}
Let us start by Taylor expanding $V$ at $p_0$ to third order,
\begin{eqnarray}
V= \Lambda_v^4 \sqrt{D}\left[v_0 + v_a\tilde{\phi}^a+ \frac{1}{2}v_{ab}\tilde{\phi}^a\tilde{\phi}^b+ \frac{1}{6}v_{abc}\tilde{\phi}^a\tilde{\phi}^b\tilde{\phi}^c\right]\,.
\end{eqnarray}
At a neighbouring point $p_1$ we have
\begin{eqnarray}
 v_0\!\mid_{p_1} &=& v_0\!\mid_{p_0} + v_a\!\mid_{p_0} \delta \tilde{\phi}^a + \frac{1}{2}v_{ab}\!\mid_{p_0} \delta \tilde{\phi}^a \delta\tilde{\phi}^b + \frac{1}{6}v_{abc}\!\mid_{p_0} \delta \tilde{\phi}^a \delta\tilde{\phi}^b \delta\tilde{\phi}^c +\dots\,, \label{taylor}\\
 v_a\!\mid_{p_1} &=& v_a\!\mid_{p_0} + v_{ab}\!\mid_{p_0} \delta \tilde{\phi}^b + \frac{1}{2}v_{abc}\!\mid_{p_0} \delta \tilde{\phi}^b \delta \tilde{\phi}^c+\dots\,, \label{taylortwo}\\
  v_{ab}\!\mid_{p_1} &=&v_{ab}\!\mid_{p_0}+v_{abc}\!\mid_{p_0} \delta \tilde{\phi}^c+\dots\,, \label{taylorthree}\\
  v_{abc}|_{p_1}&=& v_{abc}|_{p_0}+\dots\,.
\end{eqnarray}
To create a random landscape, we wish to truncate the series at third order and add a perturbation to
\begin{eqnarray}
v_{abc}\!\mid_{p_1} =v_{abc}\!\mid_{p_0}+\delta v_{abc}|_{p_0}\,,\label{perturbationvasc}
\end{eqnarray}
via an appropriately chosen $\delta v_{abc}$.
Since the third order derivative tensor, and thus $\delta v_{abc}$, has to be symmetric under permutations of $abc$, the Hessian automatically inherits the proper symmetries. As before, going along a path $\Gamma$ from point to point, we can patch together a random potential; however, this time, 
the potential, the gradient and the Hessian remain continuous. If one wishes to create a potential in $C^k$, one needs to go up to the $k+1$'th order in the Taylor expansion, as done in Sec.~\ref{Sec:potinCk}.

\subsection{Imposing properties of the Hessian}

As explained in Sec.~\ref{sec:DBMReview}, the perturbation of the Hessian needs to obey (\ref{condition1}) and (\ref{condition2}) in order to create a Dyson Brownian motion random potential.
Thus, we wish to consider $\delta v_{abc}$ 
 such that the mean and the variance of the Hessian remain the same as if perturbations were added directly to the Hessian. Consider $\delta v_{abc}$ to be Gaussian random variables. Since the sum of Gaussian random variables yields a new Gaussian random variable with mean and variances equal to the sum of the respective quantities of the summed variables, it is possible to generate a perturbation of the Hessian  $\delta v_{ab}$ with the desired properties leading to DBM.
 To be concrete, noting that
\begin{eqnarray}
 N(a,m^2) + N(b,n^2) = N(a+b, m^2 +n^2)\,, \label{additionrule} \\
 k*N(a,m^2) = N( ka , k^2m^2) \,, \label{scalarmultirule}
\end{eqnarray}
for $N(a,m^2)$ a normal distributed random variable with mean $a$ and standard deviation $m$, while $k$ is a scalar, we can work out the needed mean and variances for $\delta v_{abc}$ such that (\ref{condition1}) and (\ref{condition2}) hold, yielding
\begin{eqnarray}
 \langle \delta v_{ab}\!\mid_{p_n}\rangle &=& \langle v_{abc}\!\mid_{p_{n-1}}\delta \tilde{\phi}^c \rangle \equiv -v_{ab}\!\mid_{p_{n-1}} \frac{\|\delta \phi^a\|}{\Lambda_h}\,,\label{condition1a}\\
 \mbox{ Var}(v_{ab}|_{p_n})&=&\mbox{ Var}(v_{abc}\mid_{p_{n-1}}\delta\tilde{\phi}^c)\equiv \frac{(1+\delta_{ab})\|\delta \phi^a\|\sigma^2}{\Lambda_h} - \bigg( -v_{ab}|_{p_{n-1}}\frac{\|\delta \phi^a\|}{\Lambda_h}\bigg)^2 \,,\label{condition2a}
\end{eqnarray}
 where $\mbox{Var}(x)=<x^2>-<x>^2$. Translating the above expressions into explicit conditions for the mean and variance of $\delta v_{abc}$ is complicated by the sum over $c$. To alleviate this hurdle and simplify the procedure, we perform a rotation in field space. 

Let us consider two distinct rotations/prescriptions to identify conditions for the means and variances of $\delta v_{abc}$:
\begin{enumerate}
\item We rotate the field space such that one basis vector aligns with the step ${\boldsymbol{\delta \tilde{\phi}}}$ so that the sum collapses to a single entry.  After dividing by the step-length, we can identify the desired conditions on the means and variances. This prescription is easily generalized to generate $V \in C^k$.
\item We rotate the field space such that the Hessian is diagonalized. As a consequence, off-diagonal means are zero while variances can be simplified. 
\end{enumerate}
Once the conditions are imposed, we need to rotate back to the original coordinate system to perform the next step. While statistical properties of the Hessian are identical in both procedures, the actual conditions on $\delta v_{abc}$ differ. This is expected, since the number of independent statistical variables exceeds the number of conditions in (\ref{condition1}) and (\ref{condition2}) for $k\geq 2$. 
Since the differences are delegated to a tensor not directly entering observables, we expect all prescriptions to yield consistent predictions, e.g.~for the power-spectrum. 

\subsection{Rotating field space to align a basis vector with ${\boldsymbol{\delta \tilde{\phi}}}$ \label{Sec:rotating-align}}
We wish to rotate a basis vector $\mathbf{e}_i$ in field space in the direction of the vector  ${\boldsymbol{\delta \tilde{\phi}}}$. For $\mathbf{e}_i$ we choose the direction in which ${\boldsymbol{\delta \tilde{\phi}}}$ has its maximal component, i.e.~$|\delta \tilde{\phi}_i|>|\delta \tilde{\phi}_c|$ for $c\neq i$ (if the maximal component is degenerate, we choose the one with the lowest index).  
To perform this rotation, we identify two orthonormal vectors in the plane spanned by $\mathbf{e}_i$ and  ${\boldsymbol{\delta \tilde{\phi}}}$, which we extend to an orthonormal basis of $\mathbb{R}_D$ via the Gram-Schmidt procedure (this step is not unique). With respect to this basis, we consider the rotation by the required angle $\theta$ in the plane spanned by the two vectors of interest and use the identity on the space generated by the rest of the orthonormal basis.

Since by definition the basis vector $\mathbf{e}_i$ is already of unit norm, we begin by normalizing the vector ${\boldsymbol{\delta \tilde{\phi}}}$, 
\begin{eqnarray}
\boldsymbol{u}\equiv \frac{{\boldsymbol{\delta \tilde{\phi}}}}{|{\boldsymbol{\delta \tilde{\phi}}}|}\,.
\end{eqnarray}
To create another orthonormal vector in the plane of interest, we define
\begin{eqnarray}
\boldsymbol{v}\equiv \frac{\mathbf{e}_i - (\boldsymbol{u}\cdot\mathbf{e}_i)\boldsymbol{u}}{|\mathbf{e}_i - (\boldsymbol{u}\cdot\mathbf{e}_i)\boldsymbol{u}|}\,,
\end{eqnarray}
in line with the Gram-Schmidt method. The projection operator onto the plane spanned by $\mathbf{e}_i$ and  ${\boldsymbol{\delta \tilde{\phi}}}$ is
\begin{eqnarray}
P=\boldsymbol{u}\boldsymbol{u}^T+\boldsymbol{v}\boldsymbol{v}^T\,,
\end{eqnarray}
 and
\begin{eqnarray}
Q=I- \boldsymbol{u}\boldsymbol{u}^T - \boldsymbol{v}\boldsymbol{v}^T
\end{eqnarray}
projects onto the $D-2$ dimensional perpendicular subspace of $\mathbb{R}_D$. 
Since the rotation takes place in the target space of $P$, we can write the rotation matrix as
\begin{eqnarray}
\boldsymbol{\mathcal{R}}=I- \boldsymbol{u}\boldsymbol{u}^T - \boldsymbol{v}\boldsymbol{v}^T + [\boldsymbol{u}\ \boldsymbol{v}]R_\theta[\boldsymbol{u}\ \boldsymbol{v}]^T \label{rotationmatrix-align}
\end{eqnarray}
where $[\boldsymbol{u}\ \boldsymbol{v}]$ is the $D\times2$ matrix with $\boldsymbol{u}$ and $\boldsymbol{v}$ written as column vectors. $R_\theta$ is the normal rotation matrix in two dimensions, i.e.
\begin{eqnarray}
R_\theta = \begin{pmatrix}
\cos\theta & \sin\theta \\
-\sin\theta & \cos\theta
\end{pmatrix}\,.
\end{eqnarray}

$\boldsymbol{\mathcal{R}}$ is the desired rotation matrix aligning  $\mathbf{e}_i$ with  ${\boldsymbol{\delta \tilde{\phi}}}$ while $\boldsymbol{\mathcal{R}}^{-1}$ is used to rotate back to the original coordinate system. 
Denoting quantities in the rotated coordinate system by an underscore, we have for example
\begin{eqnarray}
\underline{v_{abc}}&=&\boldsymbol{\mathcal{R}}_{ai}\boldsymbol{\mathcal{R}}_{bj}\boldsymbol{\mathcal{R}}_{ck}v^{ijk}\,,\label{rotatetensor}\\
v_{abc}&=&\boldsymbol{\mathcal{R}}^{-1}_{ai}\boldsymbol{\mathcal{R}}^{-1}_{bj}\boldsymbol{\mathcal{R}}^{-1}_{ck}\underline{v^{ijk}}\,.\label{rotateback}
\end{eqnarray}
To avoid cluttering our notation, we suppress the underscore in the following whenever it is clear in which coordinate system computations are performed. 

\subsubsection{Imposing constraints on the mean and variance of $\delta v_{abc}$ \label{Sec:con}}
To impose (\ref{condition1a}) and (\ref{condition2a}), we go to the rotated coordinate system introduced in Sec.~\ref{Sec:rotating-align}. All expressions and tensor components in this section are given in this coordinate system (denoted by an underscore in Sec.~\ref{Sec:rotating-align}) if not stated otherwise.

We first recall that the rescaled step length is given by
\begin{eqnarray}
\|\delta\tilde{\phi}^c\|=\frac{\|\delta\phi^c\|}{\Lambda_h}=\frac{\delta s}{ \Lambda_h}\,,
\end{eqnarray}
where we keep $\delta s= \mbox{constant}$ from step to step\footnote{It is simple to incorporate a variable step length, e.g.~if one wishes to sample steep regions of the potential more sensitively, as long as $\delta s|_{p_n}\approx \delta s|_{p_{n-1}}$, by  replacing $\delta s$ in (\ref{conditionvarianceabi}) by $\delta s|_{p_{n-1}}$ . Note that the actual trajectory connecting the start and endpoint of $\boldsymbol{\delta \tilde{\phi}}$ can be curved, which is generically this case in cosmological settings where the trajectory is determined by solving the field equations given the potential in a patch. After the trajectory moved a certain distance away from the starting point (for instance the Euklidian distance $\delta s$), one defines the vector connecting the start and end point of this path's trajectory as $\boldsymbol{\delta \tilde{\phi}}$; we align this vector with one of the basis vectors in the rotated coordinate system, so that the sum $v_{abc}|_{p_{n-1}}\delta\tilde{\phi}^c$ in (\ref{condition1a}) collapses to a single term. The rotation is only needed to simplify this sum at this point in field space. The actual trajectory leading to said point does not enter.}. As  ${\boldsymbol{ \delta\tilde{\phi}}}$ and $\mathbf{e}_i$ are aligned (the index $i$ is not a free index in this section), we have
\begin{eqnarray}
{\boldsymbol{\delta \tilde{\phi}}}&=&(0,\dots,0,\delta \tilde{\phi}_i,0,\dots,0)
=\frac{\delta s}{\Lambda_h} \mathbf{e}_i\,.
\end{eqnarray}
As a consequence, (\ref{condition1a}) becomes
\begin{eqnarray}
<v_{abi}|_{p_{n-1}}>&=&-v_{ab}|_{p_{n-1}}\,.
\end{eqnarray}
Noting that $v_{abi}|_{p_{n-1}}=v_{abi}|_{p_{n-2}}+\delta v_{abi}|_{p_{n-2}}$ we arrive at the $D(D-1)/2$ independent conditions for the mean
\begin{eqnarray}
< \delta v_{abi}|_{p_{n-2}}>&=&-v_{ab}|_{p_{n-1}}-v_{abi}|_{p_{n-2}}\,.\label{conditionsmeanabi}
\end{eqnarray}
Since $v_{abc}$ is completely symmetric, the following components are determined by symmetry
\begin{eqnarray}
< \delta v_{ibc}|_{p_{n-2}}>&=&< \delta v_{cbi}|_{p_{n-2}}>\,,\\
< \delta v_{aic}|_{p_{n-2}}>&=&< \delta v_{aci}|_{p_{n-2}}>\,.
\end{eqnarray}
Since a completely symmetric tensor of rank $k$ and indices ranging from $1$ to $D$ has
\begin{eqnarray}
\mathcal{N}(k,D)=\left( \begin{array}{c}
D+k-1 \\
k 
 \end{array}\right)=\frac{(D+k-1)!}{k!(D-1)!}
\end{eqnarray} 
independent components, there are $\mathcal{N}(3,D)-\mathcal{N}(2,D)=D(D-1)^2/6$ hitherto unspecified entries in $\delta v_{abc}|_{p_{n-2}}$. These leave the conditions (\ref{condition1a}), and thus the Hessian, unaltered, but they affect the overall smoothness of the potential. Thus, imposing the GOE for the Hessian does not uniquely determine the distribution of $\delta v_{abc}|_{p_{n-1}}$ or the class of random potentials. The only constraint on these unspecified entries is that they should lead to a symmetric third order tensor. The simplest choice is to leave these components unchanged in the rotated coordinate system. As a consequence, the only variation of these entries in the original frame stems from rotating to-and-fro. This choice leads to the \lq\lq{}smoothest\rq\rq{} potential satisfying  (\ref{condition1a}).

Following a similar line of reasoning, we can use (\ref{condition2a}) to set the variance of $\delta v_{abc}|_{p_{n-2}}$. Using
\begin{eqnarray}
\mbox{Var}(v_{abc}|_{p_{n-1}}\delta\tilde\phi^c)=\left(\frac{\delta s}{\Lambda_h}\right)^2\mbox{Var}(\delta v_{abc}|_{p_{n-2}})\,,
\end{eqnarray}
we arrive at
\begin{eqnarray}
\mbox{Var}(\delta v_{abi}|_{p_{n-2}})
&=&(1+\delta_{ab})\frac{\Lambda_h}{\delta s}\sigma^2-\left(v_{ab}|_{p_{n-1}}\right)^2\ \label{conditionvarianceabi}\,,
\end{eqnarray}
where used (\ref{conditionsmeanabi}). Again, symmetries determine the values of variances related by permutations of ${a,b,i}$, while $D(D-1)^2/6$ entries are up to our choice. We again choose to leave these entries unchanged in the rotated coordinate system.

Conditions (\ref{conditionsmeanabi}) and (\ref{conditionvarianceabi}) set the mean and variance of $\delta v_{abc}$ in the {\it rotated coordinate system} $\underline{S}$ and they are deceptively simple, but it should be noted that these equations are not tensorial. Thus, they only hold in $\underline{S}$. However, since $\delta v_{abc}$ is a tensor, it can  be rotated back to the original coordinate system $S$ via (\ref{rotateback}). In $S$ the simplicity of the imposed conditions is hard to spot; indeed, working entirely in $S$, it is  hard to distinguish between conditions dictated by (\ref{condition1a}) as well as (\ref{condition2a}) and free choices made on our part.

To summarize, we arrive at totally symmetric, Gaussian perturbations $\delta v_{abc}$ such that the Hessian obeys (\ref{condition1}) and (\ref{condition2}), which in turn define a generalized Dyson Brownian motion random potential $V\in C^2$ as opposed to $V\in C^1$. A generalization to $V\in C^k$ along the same lines is straightforward, see Sec.~\ref{Sec:potinCk}.

\subsubsection{Discussion}
At this point we would like to step back and briefly compare our method to the one of Marsh et al. \cite{Marsh:2013qca}, where perturbations are added directly to the Hessian:  we first note that no additional constraint is imposed onto the step length compared to the method by Marsh et.al.: $\delta s \ll \Lambda_h$ has to hold in either case. In practice, $\delta s / \Lambda_h$ as high as 0.1 can be sufficient\footnote{there is an additional artefact for $V\in C^3$ that may require a smaller steplength, as discussed in Sec.~\ref{Sec:potinCk}; however, this artefact is not present for $V\in C^2$}.

Since the step length is the same, the computational efficiency is not improved compared to \cite{Marsh:2013qca} and in fact slightly worse,  since we need to perturb $v_{abc}$ instead of $v_{ab}$, so that more random variables need to be set at each step. Thus, computational time scales as $D^3$ (us) compared to $D^2$ (Marsh et al.). Nevertheless, either method is superior to generating potentials via a truncated Fourier series, where computational time scales as $\mbox{few}^D$.

The main improvement of our method is the following: we create a potential that can be differentiated twice ($V\in C^2$) instead of only once ($V \in C^1$). Such higher differentiability is crucial to compute cosmological perturbations, since the second derivative of the potential enters in the equations of motions (see e.g. eq. (5) in \cite{Gordon:2000hv}). In the method by Marsh, the Hessian contains jumps that lead to strong artefacts in correlation functions, see e.g.~\cite{Joy:2007na}; as a rule of thumb, the amplitude of corrections to the powerspectrum caused by a single jump scales as \cite{Joy:2007na} $\Delta \equiv [V^{\prime\prime}]_{\pm}/(3H^2)$ around scales that cross the Hubble radius at the jump of the masses (indicated by $[V^{\prime\prime}]_{\pm}$). Current observations are sensitive to such oscillatory corrections down to the percent level \cite{Easther:2013kla,Chen:2014joa} (depending on the shape and location of the feature). In DBM, the expectation value of such jumps scales with $\sim( M_p/\Lambda_h)^2 v_{ab}\delta s/(\Lambda_h v_0)\propto \delta s/\Lambda_h$ at each step. Thus, if a potential generated via DBM is to be used at the perturbed level, the step length needs to be kept sufficiently small to keep artefacts well below observational levels. Further, due to the discontinuity of the masses, one needs to apply the Deruelle-Mukhanov matching conditions \cite{Deruelle:1995kd} for perturbation in all fields at each step. In addition, one should impose a cut-off for perturbations of the Hessian if $\Delta$ approaches $0.01$ (i.e.~one needs to suppress rare outliers). Thus, using a smooth potential such as ours is advantageous, because cosmological perturbations evolve smoothly and no artefacts are generated in the first place. Of course, in the continuum limit, $\delta s/\Lambda_h\rightarrow 0$, DBM leads to a well-defined continuous stochastic Hessian as well.


\subsection{Rotating field space to diagonalize the Hessian\label{Sec:rotating-diagonalize}}

By relegating perturbations to the tensor of third derivatives, we were free to choose $D(D-1)^2/6$ entries to our liking, since they did not influence the statistical properties of the Hessian. To check whether or not this choice has strong impact on other properties of the resulting potential, we provide in this section another recipe based on rotating the field space such that the Hessian is diagonalized.  The required rotation matrix $\boldsymbol{\mathcal{R}}$ is a $D\times D$ matrix with  rows given by the eigenvectors of the Hessian, so that
\begin{eqnarray}
\boldsymbol{\mathcal{\tilde H}}= \boldsymbol{\mathcal{R}}\boldsymbol{\mathcal{H}}\boldsymbol{\mathcal{R}}^{-1}
\end{eqnarray}
 is diagonal in the rotated space. The inverse of this matrix, $\boldsymbol{\mathcal{R}}^{-1}$, is used to rotate back to the original coordinate system. The rotation to and from the rotated space is again governed by equations (\ref{rotatetensor}) and (\ref{rotateback}) respectively.
 
 \subsubsection{Imposing constraints on the mean and variance of $\delta v_{abc}$ \label{Sec:cond2}}
In this section, all the quantities are given in the rotated coordinate system of Sec.~\ref{Sec:rotating-diagonalize} unless stated otherwise and we omit the underscore notation. 
 
Before imposing (\ref{condition1a}) and (\ref{condition2a}) in the rotated coordinate system, we note that the only non-zero components of the Hessian are the diagonal ones, i.e., $v_{aa}$ (a repeated index $a$ is not to be summed over in this section). Thus, equations (\ref{condition1a}) and (\ref{condition2a}) can be re-stated as
\begin{eqnarray}
\langle \delta v_{aa}\!\mid_{p_n}\rangle = \langle v_{aac}\!\mid_{p_{n-1}}\delta \tilde{\phi}^c \rangle & \equiv & -v_{aa}\!\mid_{p_{n-1}} \frac{\|\delta \phi^a\|}{\Lambda_h}\,,\label{condition1adiagonal}\\
\mbox{ Var}(v_{aa}|_{p_n}) = \mbox{ Var}(v_{aac}\mid_{p_{n-1}}\delta\tilde{\phi}^c) & \equiv & \frac{2\|\delta \phi^a\|\sigma^2}{\Lambda_h} - \bigg( -v_{aa}|_{p_{n-1}}\frac{\|\delta \phi^a\|}{\Lambda_h}\bigg)^2 \,,\label{condition2adiagonal}
\end{eqnarray}
for the diagonal elements and 
\begin{eqnarray}
\langle \delta v_{ab}\!\mid_{p_n}\rangle = \langle v_{abc}\!\mid_{p_{n-1}}\delta \tilde{\phi}^c \rangle & \equiv & 0 \label{condition1aoffidiag}\,,\\
\mbox{ Var}(v_{ab}|_{p_n})=\mbox{ Var}(v_{abc}\mid_{p_{n-1}}\delta\tilde{\phi}^c) & \equiv & \frac{\|\delta \phi^a\|\sigma^2}{\Lambda_h} \label{condition2aoffidiag}\,,
\end{eqnarray}
for the off-diagonal elements ($a\neq b$). Again, we have the freedom to choose undetermined components of $v_{abc}$ to our liking, as long as the resulting tensor is completely symmetric under permutations of the indices $abc$. Our choice is motivated by the following wishes/observations:
\begin{itemize}
\item We wish to keep the potential as smooth as possible. Further, rotational symmetry is desirable.
\item We observe that $v_{aaa}$ enters only in the equations of the diagonal elements while other entries ($v_{aab}$) are present in the diagonal as well as off-diagonal elements and $v_{abc}$ with $a\neq b \neq c$ appear only in the off-diagonal entries \footnote{Diagonal/off-diagonal elements refers to the diagonal/off-diagonal elements of the Hessian: notice that the sum $v_{abc}\delta\tilde{\phi}^c$ does not collapse to a single term. So $v_{aac}$ with $c\neq a$ does appear in the diagonal elements of the Hessian. On the other hand $v_{aaa}$ does not appear in the off-diagonal elements of the Hessian $v_{ab}$, since $a\neq b$.}.
\end{itemize}
Diagonalizing the Hessian merely simplifies the equations sufficiently, such that the conditions (\ref{condition1adiagonal})-(\ref{condition2aoffidiag}) can be imposed. 

After some straightforward but tedious linear algebra, see appendix \ref{App:1}, we choose
\begin{eqnarray}
<v_{aaa}|_{p_{n-1}}>&=&-v_{aa}|_{p_{n-1}}\frac{\|\delta \phi^a\|}{\Lambda_h \delta \tilde{\phi}^a}\,,\\
<v_{aab}|_{p_{n-1}}>&=&0 \,,\\
<v_{abc}|_{p_{n-1}}>&=&0\,,
\end{eqnarray}
for the means. Since $v_{abi}|_{p_{n-1}}=v_{abi}|_{p_{n-2}}+\delta v_{abi}|_{p_{n-2}}$, we have
\begin{eqnarray}
<\delta v_{aaa}|_{p_{n-2}}>&=&-v_{aa}|_{p_{n-1}}\frac{\|\delta \phi^a\|}{\Lambda_h  \delta \tilde{\phi}^a} - v_{aaa}|_{p_{n-2}}\,,\label{conditionmean1}\\
<\delta v_{aab}|_{p_{n-2}}>&=& - v_{aab}|_{p_{n-2}} \,,\\
<\delta v_{abc}|_{p_{n-2}}>&=& -v_{aac}|_{p_{n-2}}\,.\label{conditionmean3}
\end{eqnarray}
For the variance, we choose
\begin{eqnarray}
\mbox{Var}(\delta v_{aaa}|_{p_{n-2}})&=&\frac{\delta s}{\Lambda_h}\frac{1}{\sum_j{\delta \tilde{\phi}^j}^2}\sigma^2 + \frac{\delta s}{\Lambda_h}\frac{1}{{\delta \tilde{\phi}^a}^2}\sigma^2 -\bigg( -v_{aa}|_{p_{n-1}}\frac{\delta s}{\Lambda_h \delta \tilde{\phi}^a}\bigg)^2 \,,\label{conditionvariance1}\\
\mbox{Var}(\delta v_{aab}|_{p_{n-2}})&=&\frac{\delta s}{\Lambda_h}\frac{1}{\sum_j{\delta \tilde{\phi}^j}^2}\sigma^2  \,,\label{conditionvariance2}\\
\mbox{Var}(\delta v_{abc}|_{p_{n-2}})&=&\frac{\delta s}{\Lambda_h}\frac{1}{\sum_j{\delta \tilde{\phi}^j}^2}\sigma^2 \,.\label{conditionvariance3}
\end{eqnarray}
Propagating these variances in accordance with (\ref{additionrule}) and (\ref{scalarmultirule}) results in (\ref{condition2adiagonal}) and (\ref{condition2aoffidiag}). The remaining elements are determined by the symmetry under permutation of indices.

This choice is straightforward, except for the variance of $v_{aaa}$, which can be read off once symmetry under rotations is imposed and everything else is fixed. 
Once the elements of the third order tensor are set in  the rotated frame, we rotate them back to the original one by  using the inverse rotation matrix $\boldsymbol{\mathcal{R}}^{-1}$.







\section{Generating random potentials $V \in C^k$ \label{Sec:potinCk}}

%
%

To generate a potential $V \in C^k$ while maintaining a Hessian in the GOE, we need to continue the Taylor expansion in (\ref{taylor}) to order $k+1$, while adding the perturbation to the $k+1$'th derivative tensor. To keep our notation economical, let us introduce the multi-index $C_j$,
\begin{eqnarray}
C_1&\equiv& c_1\,,\\
C_2&\equiv& c_1c_2\,,\\
\nonumber &\vdots& \\
C_{k-1}&\equiv& c_1\dots c_{k-1}\,,
\end{eqnarray}
so that e.g.~the third derivative tensor reads $v_{abc}=v_{abC_1}$. We kept the indices $a,b$ explicit, since we want to impose conditions (\ref{condition1}) and (\ref{condition2}) onto the Hessian $v_{ab}$. We add perturbations via
\begin{eqnarray}
v_{abC_{k-1}}|_{p_n}=v_{abC_{k-1}}|_{p_{n-1}}+\delta v_{abC_{k-1}}|_{p_{n-1}}\,,\label{perturbternsork-1}
\end{eqnarray}
staying with our convention that $\delta v_{abC_{k-1}}|_{p_{n-1}}$ is added at $p_n$ and thus relevant for the potential along $\Gamma$ from point $p_n$ onward.
 
From here on we work in the rotated coordinate system introduced in Sec.~\ref{Sec:rotating-align}, i.e.~all tensors are assumed to be transformed via (\ref{rotatetensor}) (omitting the underscore), and we assume a constant distance between perturbations so that ${\boldsymbol{\delta \tilde{\phi}}}= {\mathbf e}_i \delta s / \Lambda_h=\mbox{constant}$. Note that $i$ is not a summation index in this section but designates the direction of $\boldsymbol{\delta \tilde{\phi}}$ in the rotated coordinate system. As a consequence, we may write
\begin{eqnarray}
<\delta v_{ab}|_{p_n}>=\left<\sum_{j=1}^{k-1}\frac{1}{j!}v_{abC_j}|_{p_{n-1}}\left(\frac{\delta s}{\Lambda_h}\right)^j\right>\,. 
\end{eqnarray}
where the multi index becomes
\begin{eqnarray}
C_1&=&i\,,\\
C_2&=&ii\,,\\
\nonumber &\vdots& \\
C_{k-1}&\equiv& \underbrace{i\dots i}_{k-1}\,.
\end{eqnarray}
Since the perturbation enters only in $\delta v_{abC_{k-1}}|_{p_{n-1}}$, the remaining deterministic terms can be taken out of the expectation value to yield
\begin{eqnarray}
\nonumber <\delta v_{ab}|_{p_n}>&=&\sum_{j=1}^{k-2}\frac{1}{j!}v_{abC_j}|_{p_{n-1}}\left(\frac{\delta s}{\Lambda_h}\right)^j+\frac{1}{(k-1)!}v_{abC_{k-1}}|_{p_{n-2}}\left(\frac{\delta s}{\Lambda_h}\right)^{k-1}\\
&&+\frac{1}{(k-1)!}\left(\frac{\delta s}{\Lambda_h}\right)^{k-1}<\delta v_{abC_{k-1}}|_{p_{n-2}}>\,,
\end{eqnarray}
 where we used (\ref{perturbternsork-1}). Equating the above with the r.h.s.~of equation (\ref{condition1}) 
 yields
 \begin{eqnarray}
\nonumber <\delta v_{abC_{k-1}|_{p_{n-2}}}>&=&-(k-1)!\left(\frac{\Lambda_h}{\delta s}\right)^{k-2} v_{ab}|_{p_{n-1}}-\sum_{j=1}^{k-2}\frac{(k-1)!}{j!}v_{abC_{j}}|_{p_{n-1}}\left(\frac{\delta s}{\Lambda_h}\right)^{j+1-k}\\
 &&-v_{abC_{k-1}}|_{p_{n-2}}\,.\label{conditionsmeangeneralabi}
 \end{eqnarray}
Since $v_{abC_{k-1}}$ is completely symmetric, all means related to $<\delta v_{abC_{k-1}|_{p_{n-2}}}>$ via a permutation of the indices, $\pi(abC_{k-1})\equiv \pi(abi\dots i)$, are identical to the above expression. We leave the remaining $N(k+1,D)-N(2,D)$ elements of $v_{abC_{k-1}}$ unperturbed, i.e., impose zero mean and variance, as we did for $V\in C^2$. The above result reduces to (\ref{conditionsmeanabi}) for $k=2$. Similarly, we can derive the variance: since only $v_{abC_{k-1}}$ contains a perturbation we get
 \begin{eqnarray}
\mbox{Var}(v_{ab}|_{p_n})&=&\mbox{Var}\left(\frac{1}{(k-1)!}v_{abC_{k-1}}|_{p_{n-1}}\left(\frac{\delta s}{\Lambda_h}\right)^{k-1}\right)\,,\\
&=&\left(\frac{1}{(k-1)!}\left(\frac{\delta s}{\Lambda_h}\right)^{k-1}\right)^2\mbox{Var}\left(\delta v_{abC_{k-1}}|_{p_{n-2}}\right)\,.
 \end{eqnarray} 
Plugging the above into (\ref{condition2}) leads to
\begin{eqnarray}
\mbox{Var}\left(\delta v_{abC_{k-1}}|_{p_{n-2}}\right)=\left((k-1)!\left(\frac{\Lambda_h}{\delta s}\right)^{k-1}\right)^2\left((1+\delta_{ab})\frac{\delta s}{\Lambda_h}\sigma^2-\left(\frac{\delta s}{\Lambda_h}v_{ab}|_{p_{n-1}}\right)^2\right)\,.\label{conditionvariancegeneralabi}
\end{eqnarray} 
Variances with indices given by a permutation $\pi(abi\dots i)$ are identical to the above expression, while all remaining variances are set to zero. (\ref{conditionvariancegeneralabi}) reduces to (\ref{conditionvarianceabi}) for $k=2$.
 
Evidently, it is straightforward to set the mean and variance of $\delta v_{abC_{k-1}}$ in the rotated coordinate system such that the Hessian is a matrix in the GOE and $V\in C^{k}$. The above values need to be transformed back to our original coordinate system via (\ref{rotateback}), in which the conceptual simplicity is hidden.

\section{Examples, comparisons and discussion of $V\in C^i$ with $i=1,2,3$ \label{Sec:finaldiscussion}}
In addition to the known method of generating $V\in C^1$ via Dyson Brownian motion, see Sec.~\ref{sec:DBMReview}, we have derived two distinct methods to generate random potentials $V\in C^2$, one of which we generalized to provide $V\in C^k$ for arbitrary $k\in \mathbb{N}$; thus, we have at our disposal:
\begin{enumerate}
\item $V\in C^1$, generated via Dyson Brownian motion, see Sec.~\ref{sec:DBMReview}.
\item $V\in C^2$, generated via rotating field space to align a basis vector with $\boldsymbol{\delta \tilde{\phi}}$, Sec.~\ref{Sec:rotating-align}, yielding the conditions (\ref{conditionsmeanabi}) and (\ref{conditionvarianceabi}) for the means and variances.
\item $V\in C^2$, generated via rotating field space to diagonalize the Hessian, yielding the conditions (\ref{conditionmean1})-(\ref{conditionmean3}) and (\ref{conditionvariance1})-(\ref{conditionvariance3}) for the means and variances.
\item $V\in C^k$, generated via rotating field space to align a basis vector with $\boldsymbol{\delta \tilde{\phi}}$ and delegating perturbation to the $k+1$'th derivative tensor, yielding (\ref{conditionsmeangeneralabi}) and (\ref{conditionvariancegeneralabi}) for the means and variances.
\end{enumerate}
In this section we would like to compare the resulting potentials for $k=1,2,3$ with each other based on a few selected examples to highlight general features.

\begin{figure}[t]
\centering
\subfigure[ $V\in C^1$ - via Dyson Brownian Motion.]
{\includegraphics[width=0.45\linewidth]{Compare/potential-marsh}
 }
\subfigure[ $V\in C^2$ - Rotation to diagonalize Hessian.]
{\includegraphics[width=0.45\linewidth]{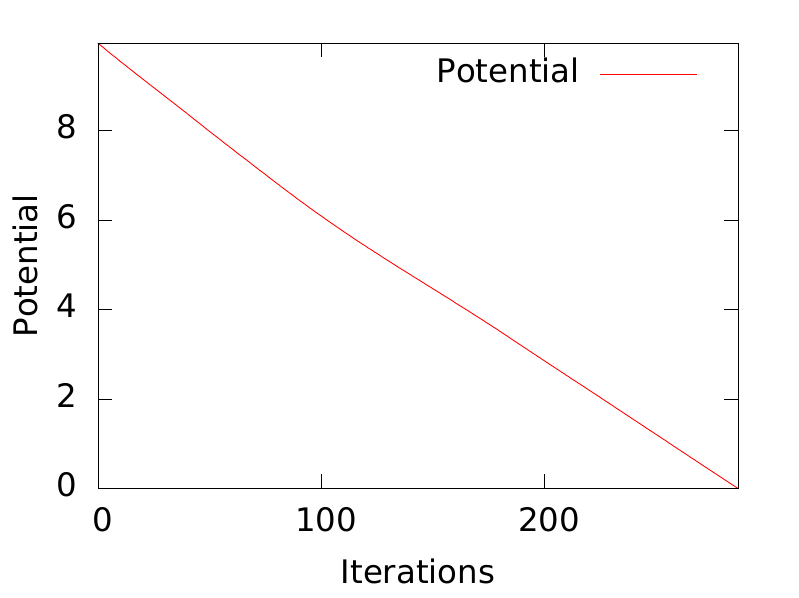}
 }
\subfigure[ $V\in C^2$ - Rotation to align basis vector with $\boldsymbol{\delta \tilde{\phi}}$.]
{\includegraphics[width=0.45\linewidth]{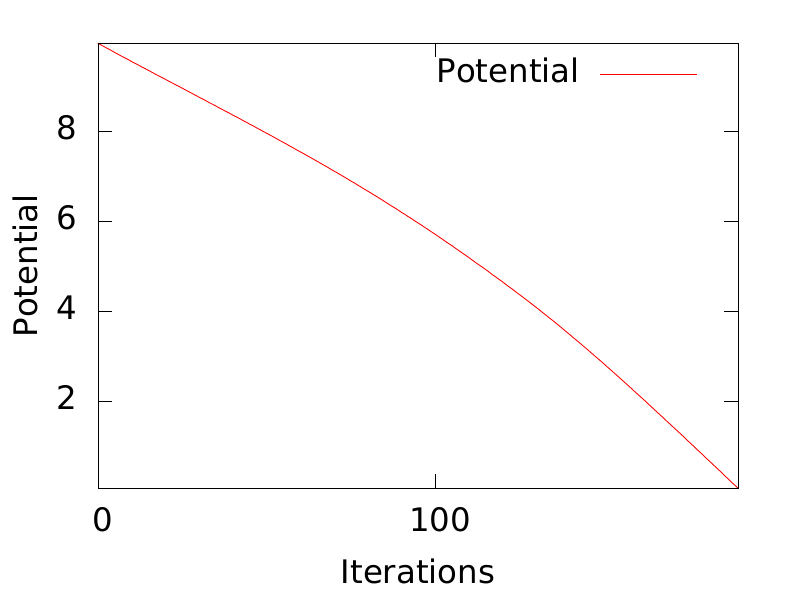}
 }
\subfigure[ $V\in C^3$.]
{\includegraphics[width=0.45\linewidth]{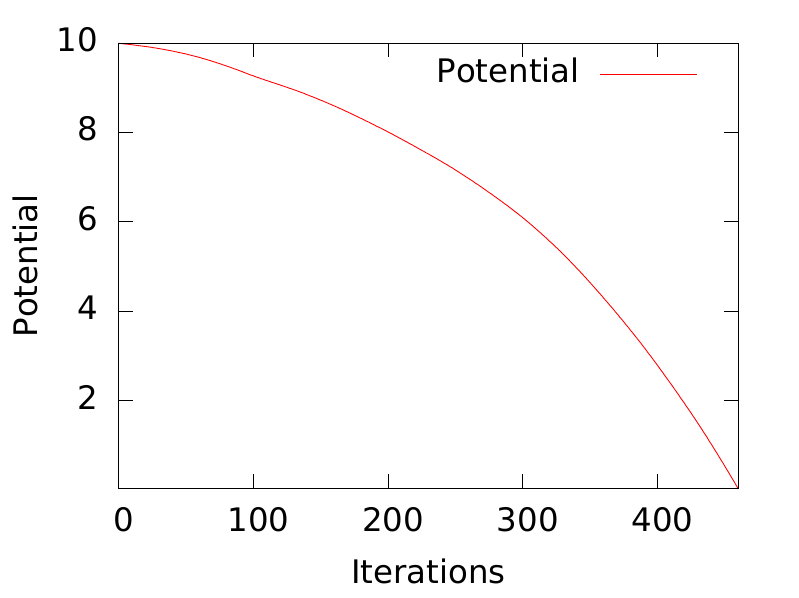}
 }
\caption{Potential along the path $\Gamma$ set by following the steepest descent. Potentials created via the methods in panels (a)-(b) are, on average, similar.  Potentials created via aligning a basis vector with $\boldsymbol{\delta \tilde{\phi}}$ are more often convex, $V\in C^3$ more so than $V\in C^2$.  \label{Fig:plotpotential}}
\end{figure}

 As explained in Sec.~\ref{sec:eigenvaluerelaxation},  we use $\Lambda_h=0.1$, $\Lambda_v=1$, $\delta s=\Lambda_h/100$, $\sigma^2=2/D$, and chose the direction of steepest descent to provide the path $\Gamma$ along which the potential is generated.  We choose $D=5$, so that we are well within the multi-field regime and avoid self intersecting $\Gamma$, yet plots, such as the ones depicting eigenvalue relaxation of the Hessian, remain clear. Further, to ease comparison, we use the same initial configuration in plots (height, slope and eigenvalues of the Hessian); we use the words ``iterations'' and ``steps'' interchangeably. We varied initial conditions and the dimensionality of field space to make sure that the plots depicted here are representative. 

In Fig.~\ref{Fig:plotpotential}, we show exemplary plots of the potential along the path $\Gamma$: we observe that $V\in C^2$ generated via either method yields potentials comparable to the one originating from Dyson Brownian motion. However, potentials $V\in C^3$, panel (d), and to a lesser degree $V\in C^2$ in panel (c), are generically more convex than the other ones. While these differences are minor, they may be important if questions pertaining to inflationary cosmology are to be addressed (only flat regions can support inflation); thus, for applications in inflationary cosmology, it is crucial to check the sensitivity of predictions to the method used to generate the potential. We leave such an investigation to future work. 

\begin{figure}
\centering
\subfigure[$V\in C^1$ - via Dyson Brownian Motion.]
{\includegraphics[width=0.45\linewidth]{Compare/eigen-marsh}
 }
\subfigure[$V\in C^2$ - Rotation to diagonalize Hessian.]
{\includegraphics[width=0.45\linewidth]{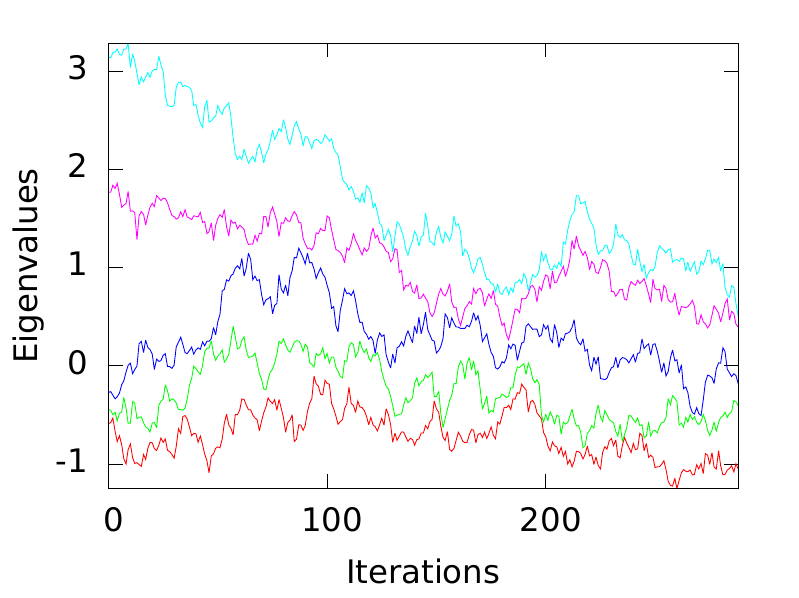}
 }
\subfigure[$V\in C^2$ - Rotation to align basis vector with $\boldsymbol{\delta \tilde{\phi}}$.]
{\includegraphics[width=0.45\linewidth]{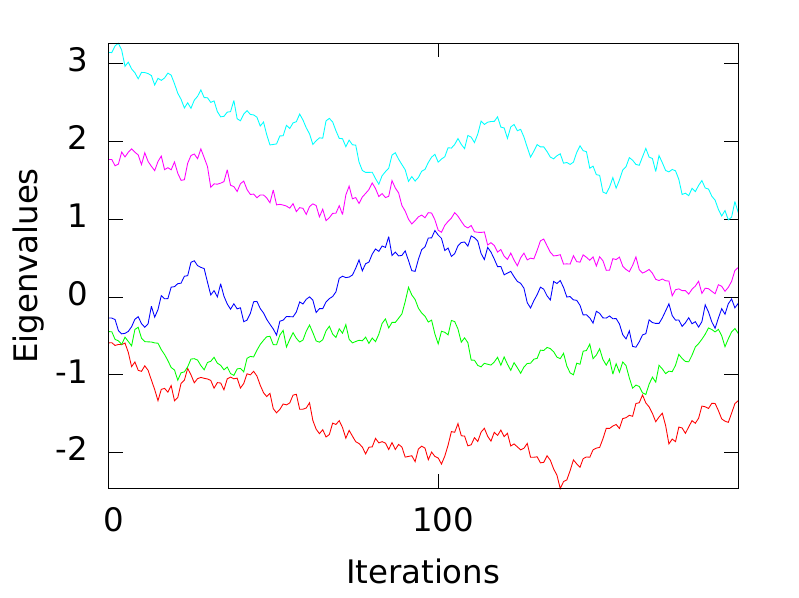}
 }
\subfigure[$V\in C^3$.]
{\includegraphics[width=0.45\linewidth]{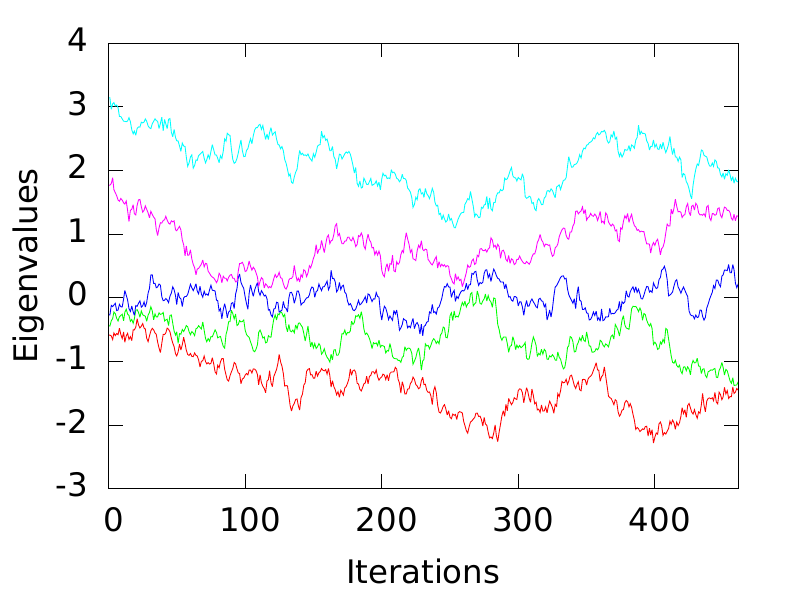}
 }
\caption{The eigenvalues of the Hessian plotted along the path $\Gamma$ for the potentials shown in Fig.~\ref{Fig:plotpotential}. All potentials show eigenvalue-relaxation, as expected for a Hessian in the Gaussian orthogonal ensemble. \label{fig:eigenvaluerelaxation}}
\end{figure}

The main desired difference between potentials is their differentiability. This difference becomes evident if we plot the eigenvalues of the Hessian along the path, as in Sec.~\ref{sec:eigenvaluerelaxation}. The corresponding eigenvalues of the potentials in Fig.~\ref{Fig:plotpotential} are shown in Fig.~\ref{fig:eigenvaluerelaxation}. As expected for potentials whose Hessian is a member of the Gaussian orthogonal ensemble for distances above the horizontal correlation length $\Lambda_h$, we observe eigenvalue relaxation once the path-length exceeds $\Lambda_h$; in other words, the initial values of the eigenvalues are usually forgotten after $\mathcal{O}(\Lambda_h/\delta s)=\mathcal{O}(100)$ iterations.

\begin{figure}
\centering
\subfigure[$V\in C^1$ - via Dyson Brownian Motion.]
{\includegraphics[width=0.45\linewidth]{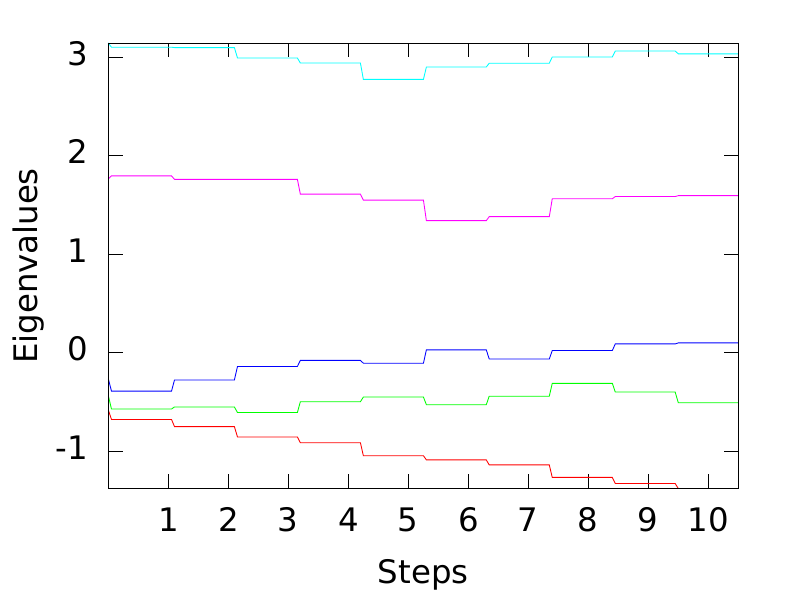}
 }
\subfigure[$V\in C^2$ - Rotation to Diagonalize Hessian]
{\includegraphics[width=0.45\linewidth]{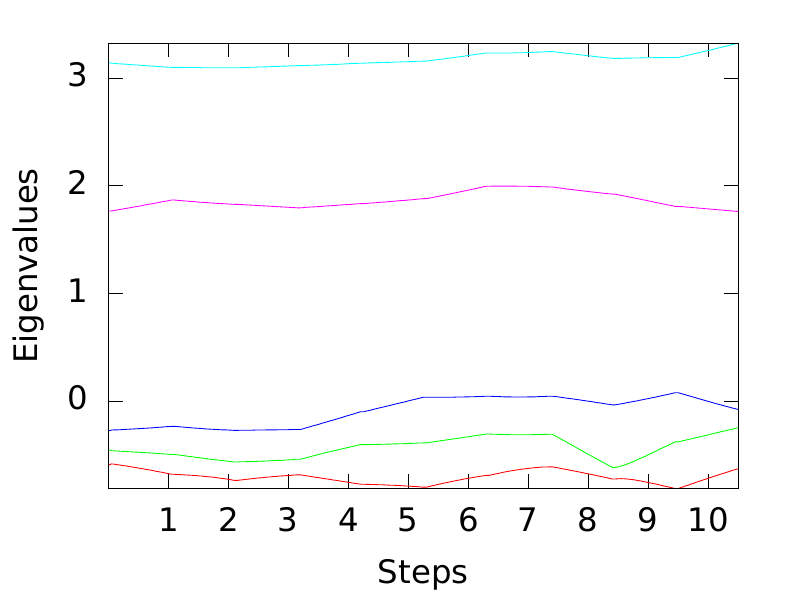}
 }
\subfigure[$V\in C^2$ - Rotation to align basis vector with $\boldsymbol{\delta \tilde{\phi}}$.]
{\includegraphics[width=0.45\linewidth]{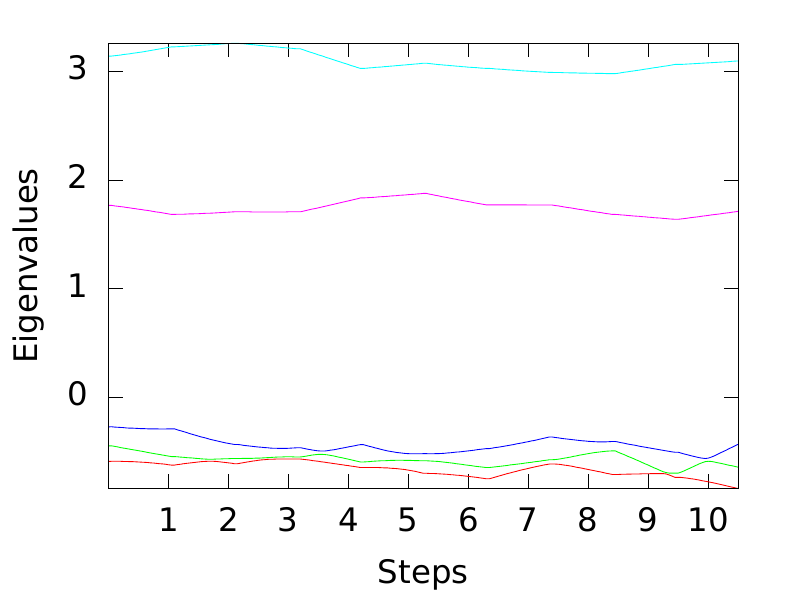}
 }
\subfigure[$V\in C^3$]
{\includegraphics[width=0.45\linewidth]{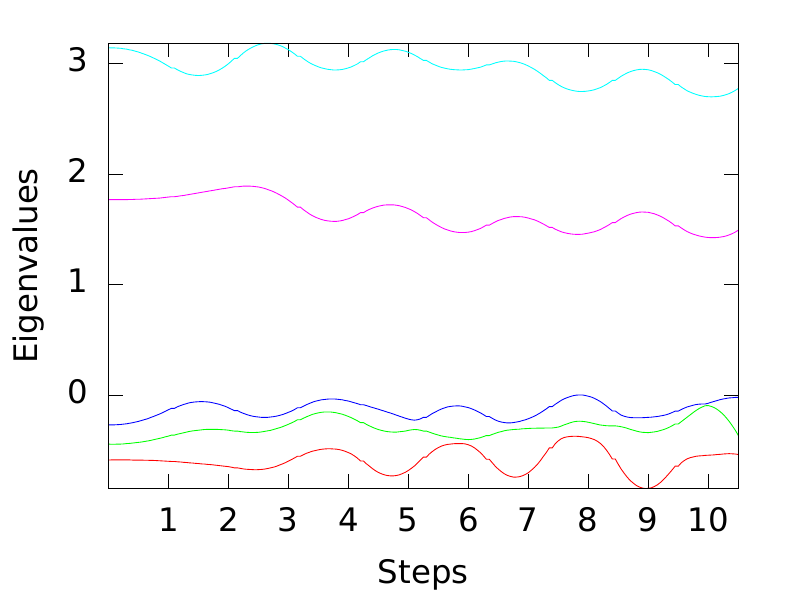}
 }
\caption{Eigenvalue evolution between successive steps for $\delta s=\Lambda_h/100$ and different methods of creating random potentials: (a) Dyson Brownian motion causes jumps at each step; (b) and (c): either method of creating $V\in C^2$ leads to a continuous evolution of the Hessian, sufficient for several cosmological applications (hunt for minima, computation of power-spectrum); (d) For $k\geq 3$ spurious oscillations arise, that can be problematic for applications. While they are intrinsic to the method we used to create such potentials, their amplitude can be made arbitrarily small by reducing $\delta s$, see Fig.~\ref{Fig:steplength}.\label{Fig:closeupeigenvalue}}
\end{figure}

While such plots look qualitatively the same to the naked eye, a close up, as shown in Fig.~\ref{Fig:closeupeigenvalue}, reveals the most important quantitative difference: potentials $V\in C^1$ show a jump of the eigenvalues after each step. These discontinuities are intrinsic to Dyson Brownian motion and, as discussed in Sec.~\ref{sec:review1}, can be disastrous if cosmological perturbations generated during inflation are to be computed (artefacts  arise, such as ringing patterns in correlation functions). Further, such potentials are not well suited to search for minima, which restricts their usability to model string theoretical landscapes if one's goal is to find suitable vacua for our universe. As we go to $V\in C^2$, panel (b) and (c) of Fig.~\ref{Fig:closeupeigenvalue}, the eigenvalues change smoothly, as expected. Since perturbations are delegated to the tensor of third derivatives, we still observe kinks, but these kinks are harmless if one wishes to compute the power-spectrum or search for minima. Nevertheless, they lead to spurious signals in higher order correlation functions, which motivated us to create potentials in even higher differentiability classes.

The eigenvalues of a potential $V\in C^3$ are plotted in  panel (d) of Fig.~\ref{Fig:closeupeigenvalue}; while the slope indeed changes continuously, as desired, we observe an artefact of a different kind: the perturbations added to the 4'th derivative tensor add spurious oscillations to the eigenvalues with a wavelength set by the step length 
\begin{eqnarray}
\lambda \propto \delta s\,.
\end{eqnarray}
 The cause of these oscillations is similar to over-fitting data with a polynomial of high degree, and we expect them to be problematic for the computation of  the bi-spectrum.

Can we eliminate this effect? Since we stitch together different Taylor expansions after each step, and these oscillations trace back to changes in the 4'th derivative tensor, one may hope  to reduce the step-length to the point where the contribution of the fourth order tensor are well below the ones of the third order tensor when the next patch is reached. For the depicted eigenvalues of the Hessian ($V\in C^3$), we stitch together parabola -- thus, ideally, the step-length should be taken so small that the presence of a maximum/minimum of the particular Taylor expansion is unlikely to occur within the next step, i.e.~$<\delta v_{abii}>  \delta s^2 \,\ll \, \mathcal{O}(v_{abi} \delta s,v_{abii} \delta s^2 )$. However, since the means of the perturbations are adjusted according to  (\ref{conditionsmeangeneralabi}), this condition can not be satisfied generically due to the contribution $<\delta v_{abii}>\,\sim \mathcal{O}(v_{abi}/\delta s)$: the left and right hand side of our tentative condition are generically of the same order. Thus, the presence of such oscillations is intrinsic to the method by which we create $V\in C^3$. 

\begin{figure}
\centering
\subfigure[$\delta s =\Lambda_h/10$]
{\includegraphics[width=0.45\linewidth]{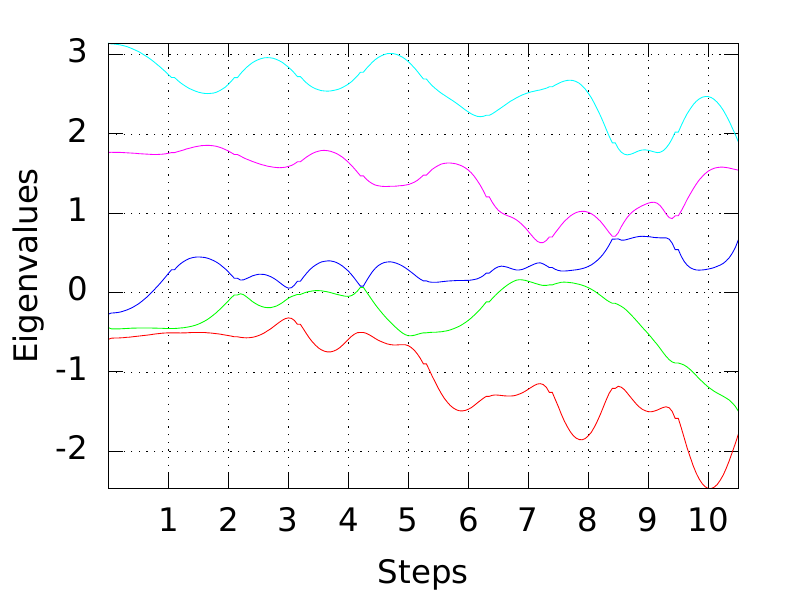}
 }
\subfigure[$\delta s =\Lambda_h/10^2$]
{\includegraphics[width=0.45\linewidth]{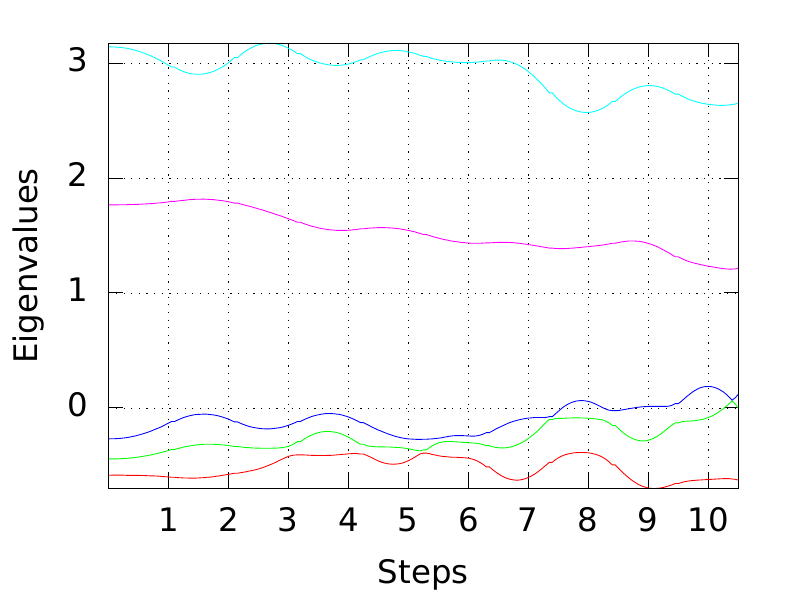}
 }
\subfigure[$\delta s =\Lambda_h/10^3$]
{\includegraphics[width=0.45\linewidth]{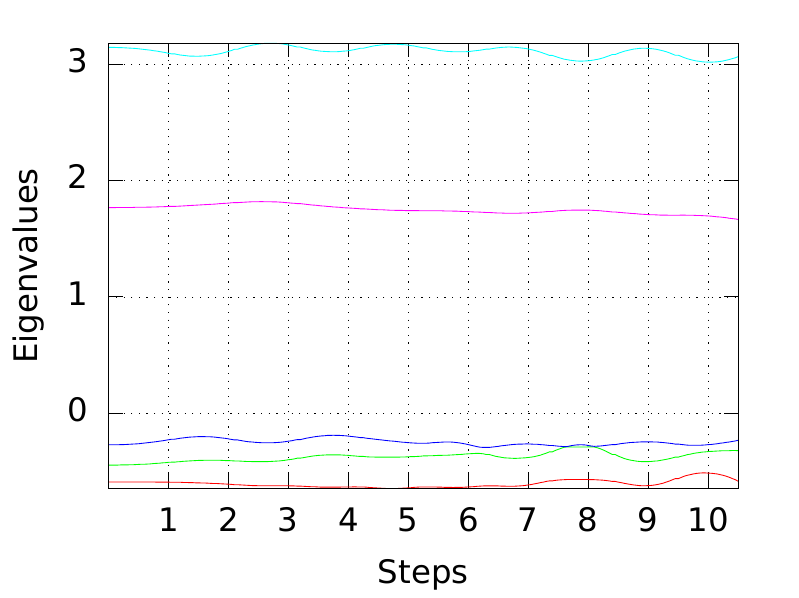}
 }
\subfigure[$\delta s =\Lambda_h/10^4$]
{\includegraphics[width=0.45\linewidth]{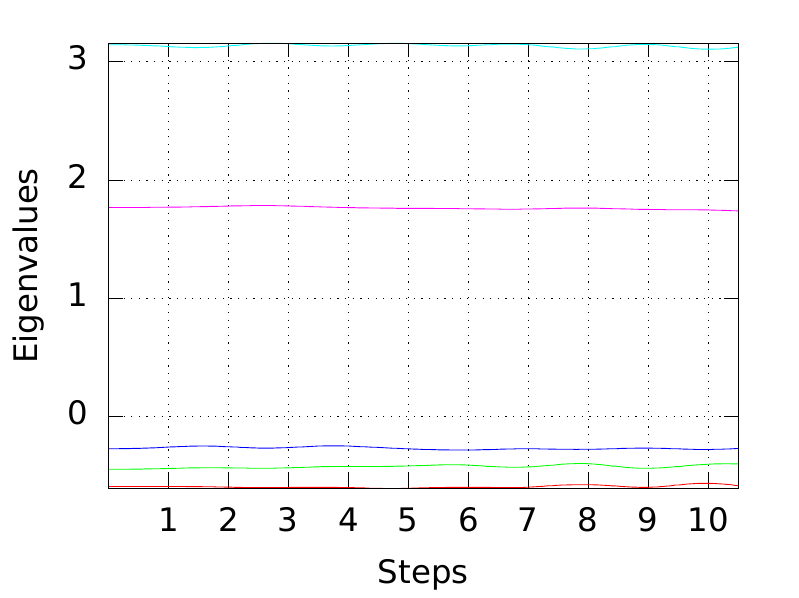}
 }
\caption{Eigenvalues for potentials $V\in C^3$ for varying step size $\delta s$. Spurious oscillations with a wavelength comparable to $\delta s$ are visible. While their presence is intrinsic to the method used to create the potential, one can diminish their amplitude to any desired level by reducing $\delta s$ accordingly.  \label{Fig:steplength}}
\end{figure}

 To test this explanation, we varied $\delta s$ from $\Lambda_h/10$ to $\Lambda_h/10^{4}$ in Fig.~\ref{Fig:steplength}, and indeed, the presence of oscillations is not altered by a reduction of $\delta s$, while the wavelength in field space is set by the step-length with a proportionality constant of order one. However, we also observe that the amplitude of oscillations diminishes as the step-length is decreased. To leading order in $\delta s$, we can estimate this amplitude via 
\begin{eqnarray}
A\sim <\delta v_{abii}>\delta s^2\sim \left\{v_{ab},v_{abi}\right\}\Lambda_h\delta s \,,
\end{eqnarray}
that is $A\propto \delta s$. The observed reduction is slightly weaker, which is understandable since our estimate did not take into consideration the variance of the perturbations in (\ref{conditionmean1})-(\ref{conditionmean3}).

However, it is clear how to proceed in applications, such as the computation of the bi-spectrum: to minimize the effect of oscillations, one needs to demand (at least) that $A$ is considerably smaller than the eigenvalues under consideration. We normalized the potential such that eigenvalues are of order one, so that we need to demand $A\ll 1$ which directly translates into a condition for $\delta s$. Practically, one may create a few plots of the eigenvalues as in Fig.~\ref{Fig:steplength} to decide on an appropriately small $\delta s$ (in our case, $\delta s\lesssim \Lambda_h/10^4$ appears appropriate). If observables such as the bi-spectrum are to be computed, one should check whether reducing  $\delta s$ further causes leading order changes in results (it should not). A similar line of reasoning needs to be followed for $k>3$.

Alternatively, one may contemplate altering the method whereby the potential is generated. For example, one may consider applying a running average to the potential as it is being created to eliminate the effect of noisy artifacts  on scales of order $\delta s$. We leave such improvements to future studies.

To summarize, while potentials $V\in C^k$ with $k\geq 3$ are not free of problems, they still offer an improvement over potentials in a lower differentiability class. In the end, one may pick the potential with the lowest $k$ that is sufficient for the task at hand in order to retain the computational advantage that locally created random potentials offer over globally created ones. Potentials $V\in C^2$ are sufficient to hunt for minima on landscapes in string theory and enable a computation of the power-spectrum of cosmological perturbations. Further, they are free of the artificial oscillations that occur for $k\geq 3$. Thus, we plan to use such potentials in forthcoming publications on cosmological applications.

\section{Conclusion}
We derived novel methods to generate random functions in a desired differentiability class along a trajectory by extending the prescription of Dyson Brownian motion (DBM). As in DBM, the Hessian of these functions evaluated at well separated points is a random matrix in the Gaussian orthogonal ensemble (GOE). 

We were motivated to construct such functions to model  complicated potentials on the string theoretical landscape (a field space of high dimensionality) for cosmological applications. Particularly potentials $V\in C^2$ are of interest to us, since they enable the search for minima as well as the study of cosmological perturbations and the computation of the power-spectrum (the two-point correlation function). Potentials $V\in C^k$ with $k\geq 3$ are needed to compute higher order correlation functions.

The method of constructing such potentials inherits the basic idea from DBM to stitch together local patches wherein the potential is given as a truncated Taylor series. Whenever the next patch is entered, random variables are added to the $k+1$'th derivative tensor ($k=1$ for DMB, so perturbations are added to the Hessian). For DBM, the statistical properties of these variables are entirely determined by the desired ensemble (the GOE) of the Hessian. However, for $k\geq 2$, additional freedom is present.

To explore this freedom, we provided two distinct prescriptions for $k=2$: the first generates potentials that invoke the least number of random variables and thus provides, in a sense, the smoothest potentials. This prescription is readily extended to arbitrary $k\in \mathbb{N}$. The second prescription perturbs primarily in the principal directions of the Hessian. It should be noted that all such potentials are indistinguishable if the statistical properties of the Hessian are used as a discriminator.

We followed with a small selection of examples to highlight the properties of potentials with $k=2,3$: for $k=2$, we found potentials generated via the two different methods to be qualitatively indistinguishable and free of artefacts; we plan to use both of them in cosmological applications in a future publication. The $k=3$ case is somewhat problematic: we observed spurious oscillations of eigenvalues with a wavelength given by the step length after which perturbations are added. These artefacts are intrinsic to the methods used, but can be made arbitrarily small by reducing the step length. While not optimal from a computational efficiency point of view, such potentials can at least in principle be used to compute higher order correlation functions.

While our motivations stem from cosmology, the method to construct such random functions is general and may be of use in other areas of science.  

\acknowledgments
We would like to thank D.~Marsch, F.G.~Pedro and A.~Westphal for discussions motivating this study, as well as D.~Battefeld, L.~Schmidt and T.~Bachlechner for comments on the draft.  C.M.~acknowledges support of the Deutsche Forschungs Gemeinschaft (DFG) covering a three month stay at the University of G\"ottingen, during which this work was instigated. C.M.~would also like to thank P.~Parmananda for valuable feedback as well as A.~Rawat and A.~Dongare for help on debugging the code and feedback. T.B.~would like to thank P.~Peter and the Institut Astrophysique de Paris (AIP) for hospitality and  support during the final stages of this work.

\appendix

\section{Mean and Variance for a diagonalized Hessian\label{Sec:appendixone} \label{App:1}}

All the quantities are given in the rotated coordinate system of Sec.~\ref{Sec:rotating-diagonalize} unless stated otherwise and we omit the underscore notation. In Sec.~\ref{Sec:cond2}, we made two observations that helped us simplify the choice of means and variances of $v_{abc}$, while satisfying the conditions (\ref{condition1a}) and (\ref{condition2a}). The conditions (\ref{condition1a}) and (\ref{condition2a})  were re-stated in (\ref{condition1adiagonal}) - (\ref{condition2aoffidiag}). Here, we mention the steps we took explicitly to avoid confusion.

The first observation (the wish for a smooth potential and rotational symmetry)  allows us to set as many variables as possible to zero while keeping the minimum number of non zero variables (which we choose on the basis of simplicity of the expression, both in form and number) in order to form a smooth Hessian and hence a smooth potential. To use the second observation ($v_{aaa}$ is only present in the diagonal elements), we explicitly write out the taylor expansion for the Hessian's elements,
\begin{eqnarray}
\delta v_{aa}|_{p_{n}}&=& v_{aaa}|_{p_{n-1}}\delta \tilde{\phi}^a + \sum_{b\neq a}{v_{aab}\delta\tilde{\phi}^b}\label{taylordiag}\\
\delta v_{ab}|_{p_{n}}&=& v_{aba}|_{p_{n-1}}\delta \tilde{\phi}^a + \sum_{c\neq a}{v_{abc}\delta\tilde{\phi}^c}\label{tayloroffdiag}
\end{eqnarray}
Thus, regarding the means, (\ref{condition1aoffidiag}) and (\ref{tayloroffdiag}) along with our wish for a smooth Hessian make the choice of zero mean for $v_{aab}$ and $v_{abc}$ obvious. This in turn sets the value of the mean for $v_{aaa}$ via (\ref{condition1adiagonal}). Since $v_{abi}|_{p_{n-1}}=v_{abi}|_{p_{n-2}}+\delta v_{abi}|_{p_{n-2}}$, we have
\begin{eqnarray}
<\delta v_{aaa}|_{p_{n-2}}>&=&-v_{aa}|_{p_{n-1}}\frac{\|\delta \phi^a\|}{\Lambda_h  \delta \tilde{\phi}^a} - v_{aaa}|_{p_{n-2}}\,,\\
<\delta v_{aab}|_{p_{n-2}}>&=& - v_{aab}|_{p_{n-2}} \,,\\
<\delta v_{abc}|_{p_{n-2}}>&=& -v_{aac}|_{p_{n-2}}\,.
\end{eqnarray}

Regarding the variance, we observe that the off-diaganal components of the tensor are quite simple again. After a bit of thinking, using $\sum_j{({\delta\tilde{\phi}^j}^2/\sum_i{{\delta\tilde{\phi}^i}^2})}=1$, we arrive at (\ref{conditionvariance2}) and (\ref{conditionvariance3}) as viable choices. The variance of the diagonal elements is then the only free variable in the equation for the variance of the diagonal Hessian element and can thus be solved for, leading to
\begin{eqnarray}
\mbox{Var}(\delta v_{aaa}|_{p_{n-2}})&=&\frac{\delta s}{\Lambda_h}\frac{1}{\sum_j{\delta \tilde{\phi}^j}^2}\sigma^2 + \frac{\delta s}{\Lambda_h}\frac{1}{{\delta \tilde{\phi}^a}^2}\sigma^2 -\bigg( -v_{aa}|_{p_{n-1}}\frac{\delta s}{\Lambda_h \delta \tilde{\phi}^a}\bigg)^2 \,.
\end{eqnarray}
The remaining elements are determined by the symmetry under permutation of indices.

-----------------------------------------------------------------

\end{document}